\documentstyle[12pt]{article}
\newcommand{\dbar}{d \hspace{-.65ex}\raisebox{.8ex}{--}}
\newcommand{\smfrac}[2]{{\textstyle \frac{#1}{#2}}}

\newcommand{\pib}[1]{\left[ #1 \right]}%put in braces
\newcommand{\pil}[1]{\left\langle #1 \right\rangle} % put in <  >
\newcommand{\ov}[1]{\overline {#1}}
\renewcommand{\u}[1]{{\underline {#1}}}

\newcommand{\wsep}[2]{\hspace{#1 in} \mbox{#2} \hspace{#1in}}
\newcommand{\wsub}[1]{_{\mbox {{\scriptsize #1}}}}
\newcommand{\w}[1]{\mbox{#1}}

\newcommand{\eqnb}{\begin{equation}}
\newcommand{\eqne}{\end{equation}}

\newcommand{\lsm}{L$\sigma$M}

\newcommand{\question}[1]{}
\title{Comments on compositeness in the SU(2) linear $\sigma$ model} 
\author {M.\ D.\ Scadron$^*$ \\
  Department of Physics,\\ University of Tasmania,\\
  Hobart, Australia, 7005}
\date{}
\begin{document}
\setlength{\textheight}{130 ex}
\addtolength{\baselineskip}{1 ex}
\addtolength{\parskip}{1 ex}
\setlength{\topmargin}{-10 ex}
\maketitle
\vfill
\begin{center}\bf{Abstract}\end{center}

First we summarize the quark-level linear $\sigma$ model compositeness 
conditions and verify that indeed $m_\sigma = 2 m_q$ when $m_\pi = 0$
and $N_c=3$, rather 
than in the $N_c\rightarrow\infty$ limit, as is sometimes suggested. 
Later we show that this compositeness picture also predicts a chiral
symmetry restoration temperature $T_c = 2f_\pi$, where $f_\pi$ is the 
pion decay constant. We contrast this self-consistent $Z=0$ compositeness
analysis with prior studies of the compositeness problem.

\vfill
\noindent
PACS \# 11.15.Pg, 11.30.Rd To appear in PRD 
\vfill
\noindent{\small $^*$ Permanent address:
  Physics Department, University of Arizona, Tucson, AZ 85721, USA }
\newpage
	Now that the scalar $\sigma$ meson has been reinstated in the
1996 particle data group tables [1], it is appropriate to take seriously
the various theoretical implications of a quark-level linear $\sigma$
model (\lsm ) field theory.  The original spontaneously broken \lsm\ 
theory [2] was recently dynamically generated [3] at the quark level
in the spirit of Nambu-Jona-Lasinio [4]. In this note we summarize the 
color number $N_c$ and compositeness properties of
the above SU(2) quark-level \lsm\  and comment on the recent
\lsm\ analysis of compositeness given by Lurie and Tupper [5].

	First we  display the interacting part of the standard 
\lsm\ [2] (quark-level) lagrangian density shifted around 
the true vacuum $\pil{\vec{\pi}} = \pil{\sigma} = 0:$ 
$$
  {\cal L}\wsub{int} = g \ov{\psi} (\sigma + i\gamma_5 \vec{\tau} \cdot 
  \vec{\pi}) \psi 
  + g' (\sigma^2 + \vec{\pi}^2) \sigma 
  - (\lambda / 4) (\sigma^2 + \vec{\pi}^2)^2
  - f_\pi g \ov{\psi} \psi ,
\eqno{(1a)}
$$
with (spontaneously broken) chiral couplings
$$
  g = m_q / f_\pi ~~,\hspace{1in} g' = m_\sigma^2 / 2f_\pi = \lambda f_\pi ~~.
\eqno{(1b)}
$$
Once the \lsm\ scalar field is shifted to $\pil{\sigma} = 0$, 
giving rise to the interacting but 
chiral-broken \lsm\ lagrangian (1), the Lee null-tadpole condition [6] 
depicted in fig.\ 1 must be valid.  Following ref.\ [3] which exploits 
the dimensional regularization [7] characterization of these quadratic 
divergent tadpole graphs in fig.\ 1 as $\int d^4 p (p^2 - m^2)^{-1} \sim
m^2$,
 one expresses the Lee condition as
$$
  0 
  = -4 m_q
  N_f N_c
  g \cdot m_q^2
  + 0
  + 3g' \cdot m_\sigma^2,
\eqno{(2a)}
$$
where the zero on the rhs of (2a) corresponds to $m_\pi^2 = 0$ in the
chiral 
limit.  Upon using eqs.\ (1b), this Lee null-tadpole condition (2a)
becomes
$$
  \smfrac{1}{2} N_f
  N_c (2m_q)^4
  = 3 m_\sigma^4 ~~.
\eqno{(2b)}
$$
Clearly if the NJL relation [4]   
$$
m_\sigma = 2 m_q\eqno(3)
$$
is valid, then (2b) requires
$$
  N_f N_c = 6,
\eqno{(4)}
$$
or $N_c = 3$ when $N_f = 2$, the latter being an input in the SU(2) \lsm .  

	It is well known that for $\pi^o\to 2\gamma$ decay, the $N_f = 2$ 
quark triangle empirically suggests $N_c=3$ (also a \lsm\  result).
Moreover eq.(4) also follows from ``anomaly matching'' [8,9].
However we shall not invoke here the stronger (but consistent)
constraints due to dynamically generating the (quark-level) 
\lsm\, as they follow from comparing quadratic and 
logarithmically divergent integrals using (compatible) regularization
schemes [3].

	Thus the condition (4) depends on the NJL relation (3) being also
true in the \lsm .
The latter assertion follows when one dynamically generates [3] the entire 
\lsm\ lagrangian (1) starting from a simpler chiral quark model (CQM) 
lagrangian , as well as dynamically generating the two additional
equations
$$
  m_\sigma = 2 m_q, \qquad g = 2\pi/\sqrt{N_c}.
\eqno{(5)}
$$
For $N_c=3$, the latter pion-quark coupling in (5) is
$g= 2 \pi / \sqrt 3 \approx 3.63$, near the anticipated value found
from the $\pi NN$ coupling $g_{\pi NN} \sim 13.4$ so 
that $g\approx g _{\pi NN} \ / 3 g_A \sim 3.5$. Then the nonstrange
constituent quark
mass is $m_q = f _\pi 2 \pi / \sqrt 3 \approx 326$ MeV, near $M_N/3$ as
expected. But rather than repeating ref.\ [3] in detail, we 
offer an easier derivation of $m_\sigma = 2 m_q$ following only from the 
quark loops induced by the CQM lagrangian. This naturally leads to
the notion of ``compositeness''.

	To this end, we invoke the log-divergent gap equation from fig.\ 2
$$
  1 = -i4 \smfrac{1}{2}N_f
  N_c g^2
  \int \dbar^4 p
  (p^2 - m_q^2)^{-2} ~~,
\eqno{(6)}
$$
where $\dbar^4 p = (2\pi)^{-4} d^4 p$.  Equation (6) is the
chiral-limiting one-loop nonperturbative 
expression of the pion decay constant $f_\pi = m_q / g$ with the  
quark mass $m_q$ cancelling out.   This L$\sigma M$ log-divergent gap
equation (6) also holds in the context of the four-quark NJL model
[10]. Then the
one-loop-order $g_{\sigma \pi \pi}$ 
coupling depicted in fig.\ 3 is
$$
  g_{\sigma \pi \pi} = 2gm_q
  \pib{-i4 \smfrac{1}{2}N_f
  N_c g^2
  \int \dbar^4 p (p^2 - m^2_q)^{-2}}
  = 2gm_q ~~.
\eqno{(7)}
$$
The one-loop $g_{\sigma \pi \pi}$ in (7) ``shrinks'' to the 
tree-order meson-meson coupling in (1b), $g' = m_\sigma^2/ 2f_\pi$, only 
if $m_\sigma = 2 m_q$ is valid along with the GTR $f_\pi g = m_q$.  This is 
a $Z = 0$ compositeness condition [11], stating that the loosely 
bound $\sigma$ meson could be treated either as a $\ov{q}q$ bound 
state (as in the NJL picture) or as an elementary particle as in 
the \lsm\ framework of fig.3. But in either case $m_\sigma = 2 m_q$ must
hold
and therefore the additional \lsm\ Lee condition (2) requires $N_c = 3$ 
when $N_f = 2$ in (4).

	It is also possible to appreciate the one-loop order $Z=0$ 
compositeness condition in the context of the \lsm\ [3] in a different
manner.
Our version of the $Z=0$ compositeness condition
is that the log-divergent gap equation (6) can be expressed in terms of 
a four-dimensional UV cutoff as
$$
  1 =\ln(1+\Lambda^{2}/m^{2}_q)-(1+m^{2}_{q}/\Lambda^{2})^{-1},
\eqno{(8)}
$$
where we have substituted only $g = 2\pi / \sqrt N_c$ and $N_f =2$
into (6) in order to deduce (8). The numerical solution of (8)
is the dimensionless ratio $\Lambda /m_q \approx 2.3$, which
is slightly {\em larger} than the NJL ratio in (3) or in (5), 
$m_\sigma/m_q =2$. Introducing the above dynamically generated
quark mass of 326 MeV, the UV cutoff inferred from (8) (i.e. from
(6)) is $\Lambda\approx 2.3 m_q \approx 750$ MeV. This 750 MeV
cutoff in turn suggests (in the \lsm ) that 
lighter masses signal elementary particles, such as $m_\pi = 0$, 
$m{_q}\approx 325$ MeV, $m_{\sigma}=2m{_q}\approx$ 650 MeV. Heavier meson
masses than 750 MeV signal $\bar{q}q$ bound states, such as $\rho (770), 
\omega$(783), $A_1(1260),$ etc. This is the essence of the $Z = 0$
compositeness conditions of refs. [11].

Given the above eqs. (3)-(8), we are now prepared to comment in detail
on the L$\sigma M$ compositeness analysis of ref. [5]. 
Again using the log-divergent cutoff condition (8), the L$\sigma$M
renormalization 
constant $Z_{3}$ computed in eq. (3) of ref.\ [5] can be expressed as
\question{On page 4 of the old version, near the bottom left-hand
  corner of the page, you wrote $T_c^2 / 4 = f_\pi^2$.  Did you want
  me to include this somewhere in the paper.  If you did, I don't know
  where.  Please ignore this if you didn't.  Otherwise, I'll make whatever
  change you request.
} 
$$
Z_{3} = 1- {N_{c}g^{2} \over 4\pi^{2}}.\
\eqno{(9)}
$$
Then the dynamically generated \lsm\ meson-quark coupling in (5) 
indeed corresponds to $Z_{3}=0$ from (9), as anticipated. 

However the 
renormalization constant $Z_{4}$ in ref. [5] then becomes using (8),
$$
Z_{4}=1+\left[3\lambda -\frac{{2N}_{c}g^{4}}{\lambda}\right] 
\frac{1}{4\pi^{2}}. \eqno{(10)}
$$
	Ignoring for the moment the second term in (10) proportional to 
$3\lambda$, we note that the log-divergent gap equation (6) requires the 
$\pi\pi \rightarrow\pi\pi$ quark box (dynamically generated by the CQM 
lagrangian ) to ``shrink'' (as in eq.\ (7) and in fig.\ 3) to a point
contact 
term $\lambda$ provided 
that [3]
$$
\lambda = 2g^{2}.
\eqno{(11)}
$$
Equation (11) also follows from both \lsm\ couplings [2] in (1b) combined
with $g_{\sigma \pi \pi} = 2gm_q$ from (7).
Substituting (11) into the third (quark loop) term in (10), one finds
$$
Z_{4}= 1+0 - \frac{N_{c}g^{2}}{ 4{\pi}^{2}},
\eqno{(12)}
$$
(where the middle zero term in (12) corresponds to the neglected meson 
loop in contrast to ref. [5]).  Equation (12) parallels the $Z_{3}$
renormalization constant in 
(9). In these two cases
$$
Z_{3}=Z_{4}=1 -\frac{N_{c}g^{2}}{ 4{\pi}^{2}} ~~,
\eqno{(13)}
$$
and then the resulting compositeness conditions $Z_3 = Z_4 = 0$ {\em both} 
reconfirm that $g=2\pi/\sqrt{N}_{c}$, as earlier 
dynamical generated in eqs.(5).

The reason why one must neglect the second meson loop term proportional 
to $3\lambda$ in (10) is because e.g. 
$\pi_{\alpha}\pi_{\beta}\rightarrow \pi_{\gamma} \pi_{\delta}$ scattering has 
tree level (or one-loop) graphs which must {\em vanish} in the 
strict zero momentum chiral limit.  This fact was emphasized on pp 
324-327 of the text by de Alfaro et al. [DFFR] in ref. [2]. 
Specifically the quartic \lsm\ contact term $-\lambda$ is cancelled by
the cubic $\sigma$ pole term $2 {g'}^2 / m_\sigma^2 \to \lambda$ by 
virtue of the Gell-Mann-L\'evy \lsm\ meson chiral couplings in (1b).  
After the 
(tree-level) lead term cancellation between contact term $\lambda$ and 
$s,t,u, \sigma$ meson poles in the \lsm , DFFR obtain the amplitude
$$
T_{\pi \pi} \propto \frac{1}{f^{2}_{\pi}}(s 
\delta_{\alpha \beta} \delta_{\gamma\delta}+ t \delta_{\alpha \gamma}
\delta_{\beta\delta} + u \delta_{\alpha\delta}\delta_{\beta\gamma}).
\eqno{(14)}
$$  
Then DFFR in [2] note that (14) above is just the Weinberg $\pi\pi$ 
amplitude [12] when $m_{\pi}^{2}= 0$, found instead via the 
model-independent 
current algebra and PCAC rather than from the linear $\sigma$ model 
(\lsm ).  Also note that (14) indeed vanishes in the strict zero momentum
chiral limit. 
A similar chiral cancellation of the $3 \lambda$ term in (10) also holds 
in one-loop order.

	When computing the one-loop order renormalization constant $Z_4$
as done by ref.\ [5] leading to eq.\ (10) above, one must be careful to
(a) account for the DFFR-cancellation due to the soft chiral symmetry
relation $2{g'}^2 / m_\sigma^2 \to \lambda$~, (b) reorganize the perturbation
theory using the log-divergent gap equation (6) shrink quark loops
to a contact meson term $\lambda$ with $\lambda = 2g^2$ as found in 
(11).
Then even in one-loop order one must recover the Weinberg form for $\pi \pi$
scattering eq.\ (14) in a model-independent fashion.

	This means that the meson loop graph with quartic couplings 
proportional to $3\lambda^2$ contributing to $\lambda Z_4$ as
$3 \lambda^2 / 4\pi^2$ in (10) will be cancelled by fermion box graphs
which are
of higher loop order.  Although our nonperturbative approach
mixes perturbation theory loops of different order, both DFFR and our use
of the Gell-Mann-L\'evy chiral symmetry meson relation $2{g'}^2 / 
m_\sigma^2 \to \lambda$ has the bonus of our nonperturbative approach 
retaining the consistent chiral symmetry compositeness condition $Z_3 =
Z_4 = 0$ from (13).

Keeping instead the middle term in (10) proportional to $3\lambda$, ref.
[5] 
concludes that the resulting $Z_{4}=0$ (then different) compositeness
condition requires that 
the NJL limit $m_{\sigma}\rightarrow 2m_{q}$ is recovered only when 
$N_{c}\rightarrow\infty$. References [13] reach the same conclusion 
although they are not working with SU(2) chiral mesons $(\sigma,
\vec{\pi})$. In 
our opinion however, the chiral SU(2) \lsm\ (1) already has $N_{c}=3$ and
not 
$N_{c}\rightarrow\infty$ built in via the Lee condition in eqs. (2) 
but only when 
$m_{\sigma} =  2m_{q}$ in the chiral limit.  We obtain these satisfying 
results only by cancelling the middle $3\lambda$ meson term in (10)
against higher quark loop graphs. Ref.[5] does not account for the 
above DFFR cancellation.

	Finally we extend the above zero temperature $(T = 0)$ chiral
symmetry absence of quartic meson loops in eqs.\ (10), (12), (14) to
finite temperature.  Again following ref.\ [5] we write the tadpole equation
in mean field approximation at high temperatures for the quark-level SU(2)
\lsm\ as 
$$
  v \pib{ (3 + N_f^2 - 1)
	  \lambda T^2 / 12 +
	  N_f N_c g^2 T^2 / 12 +
	  \lambda (v^2 - f_\pi^2)
    } = 0
\eqno(15)
$$
for flavor $N_f = 2$ and $v = v(T)$ with $v(0) = f_\pi \sim 90 \w{MeV}$ in
the chiral limit.  The first two terms in (15) represent quartic $\sigma$
and $\vec{\pi}$ loops, while the third term involving $N_c$ is the $u$
and $d$ quark bubble loop.  The temperature factors of $T^2 / 12$ in (15)
were originally obtained from finite temperature field theory Feynman rules
[14].

	Now in fact there should be {\em no} quartic meson loop contributions
surviving in (15) due to the above DFFR-type argument or the resulting 
Weinberg $\pi \pi$ amplitude in (14), even at finite temperatures.  So
the nontrivial solution of (15) at the chiral symmetry restoration
temperature $T_c$ (where $v(T_c) = 0$) is for $N_f = 2$, $N_c = 3$ 
and $\lambda = 2g^2$, with the first two meson loop terms in (15) 
proportional to ($3+N^2_f -1)\lambda$ consequently omitted,
$$
  T_c = 2f_\pi \sim 180 \w{MeV}~~.
\eqno(16)
$$
While this predicted temperature scale in (16) had been obtained earlier
[15,16], ref.\ [5] also noted (16) above but rejected it because of the 
meson loop contributions in (15).  

We in turn claim that the first two
$\sigma$ and $\vec{\pi}$ loop terms in (15) (and the middle term in (10)
proportional to $3\lambda$) are all zero due to chiral cancellations as in
DFFR [2].  Then (15) reduces to the nontrivial solution $N_c g^2 T_c^2 / 6
 = \lambda f_\pi^2$, (leading to $T_c = 2f_\pi$) or to a quark box loop
shrinking to a meson-meson quartic
point [3] due to the log-divergent gap equation (6), itself a version
of the $Z = 0$ compositeness condition.

	Although we concur with ref.\ [5]'s choice of the finite 
temperature quark bubble sign in eq.\ (15) (as opposed to the studies
in ref.\ [15]), there is an easier way to deduce $T_c = 2f_\pi$
by studying
the single fermion loop propagator dynamically generating the quark mass [3].
Then, with no sign ambiguity arising at finite temperature one finds [17]
$$
  m_q (T) = m_q + \frac{8 N_c g^2 m_q}
		       {-m_{\sigma}^2}
		  \frac{T^2}
		       {24}~~,
\eqno(17)
$$
where the $-m_\sigma^2$ factor in (17) indicates the $\sigma$ meson
tadpole propagator generating the quark mass.  When $T = T_c$ the quark
mass
``melts'', $m_q (T_c) = 0$, and (17) reduces to
$$
  m_{\sigma}^2 = g^2 T_c^2
  \wsep{.2}{or}
  T_c = 2 f_\pi
\eqno(18)
$$
{\em provided} that $N_c = 3$ and $m_\sigma = 2m_q = 2 f_\pi g$~~.

	We believe it significant that recent numerical simulations of 
lattice gauge theories find [18] $T_c = 150 \pm 30 \w{MeV}$, consistent
with (16) and (18).  In fact the zero temperature quark-level \lsm\ theory
in ref.\ [3] is likewise compatible with the reinstated scalar $\sigma$
in the PDG tables [1] or in ref.\ [19], the latter deducing a
broad nonstrange $\sigma$ scalar as $f_0$ (400--900) with mean mass
$m_\sigma \approx 650~\w{MeV}$.  This latter scale is in fact predicted in
ref.\ [3] as $m_\sigma = 2f_\pi \frac{2 \pi}{\sqrt{3}} \approx 650~\w{MeV}$.

	Rather than starting at $T = 0$, an alternative approach to
generating a realistic low energy chiral field theory begins at the
chiral restoration temperature (with $m_q (T_c) = 0$) involving bosons
$\vec{\pi}$ and $\sigma$ alone [20] and later adds in the fundamental
meson-quark interaction in (1).  Only then does one deduce the
quark-level linear $\sigma$ model (\lsm ) field theory [21].  While issues
of $N_c = 3$ and compositeness are then postponed, the resulting \lsm\
theory in ref.\ [21] starting at $T = T_c \sim 200 MeV$ with $\lambda
\sim 20$ 
appears quite similar to the
$T = 0$ \lsm\ field theory in refs.\ [2,3] with $\lambda \approx
26$ from (11) and $T_c \approx 180 MeV$ from (16).  In effect,
what goes around comes around \ldots ~~.
\medskip
\section*{{\normalsize {\bf Acknowledgements:}}}  
This research was partially supported by the 
Australian Research Council. M.D.S. appreciates hospitality of the 
University of Western Ontario and the University of Tasmania.  He also is 
grateful to  V.\ Elias, D. \ McKeon, R.\ Mendel, V.\ Miransky and 
especially R.\ Delbourgo for insightful comments. 
\medskip
\section*{{\normalsize {\bf Figure Captions}}}
\begin{description}
\item{Fig.\ 1}  Quark and meson tadpole loops summing to zero.

\item{Fig.\ 2} Quark loops for the axial current matrix element
  $\langle 0 | A_\mu | \pi \rangle ~~$.

\item{Fig.\ 3}  Chiral quark model loops for $\sigma \to \pi \pi$.

\end{description}
\newpage
\section*{{\normalsize {\bf References}}}
\begin{enumerate}
\item %1
Particle Data Group, R.\ M.\ Barnett et al., Phys.\ Rev.\ D\u{54}, 1 (1996).

\item %2
M.\ Gell-Mann and M.\ L\'evy, Nuovo Cimento \u{16}, 705 (1960); also see V.\ de Alfaro, S.\ Fubini, G.\ Furlan and C.\ Rossetti, Chap.\ 5 in ``Currents in Hadron Physics'' (North Holland, Amsterdam, 1973).

\item %3
R.\ Delbourgo and M.\ D.\ Scadron, Mod.\ Phys.\ Lett.\ A\u{10}, 251 (1995)
regularize the \lsm\ using dimensional regularization. This was recently
extended to analytic and Pauli-Villars regularization by R.\ Delbourgo, 
A.\ Rawlinson and M.\ Scadron, submitted for publication.

\item %4
Y.\ Nambu and G.\ Jona-Lasinio, Phys.\ Rev.\ \u{122}, 345 (1961).

\item %5
D.\ Lurie  and G.\ B.\ Tupper, Phys.\ Rev.\ D \u{47}, 3580 (1993).

\item %6
B.W.\ Lee, {\em Chiral Dynamics} (Gordon and Breach, 1972) p. 12.

\item %7
See e.g.\ G.\ 'tHooft and M.\ Veltman, Nucl.\ Phys.\ B\u{44}, 189 (1972) and
	review by R.\ Delbourgo, Repts.\ Prog.\ Phys.\ \u{39}, 345 (1976).

\item %8
S.\ Adler, Phys.\ Rev.\ \u{177}, 2426 (1969); J.\ Bell 
and R.\ Jackiw, Nuovo Cimento \u{60}, 47 (1969). See also J.\ Schwinger,
Phys. Rev. \u{82}, 664 (1951).

\item %9
G.\ 't Hooft, in Recent developments in gauge theories, 
ed.\ G.\ 't Hooft et al. (Plenum, NY 1980); Y.\ 
Frishman, A.\ Schwimmer, T.\ Banks and S.\ Yankielowicz, 
Nucl.\ Phys.\ B\u{177}, 157 (1981); S.\ Coleman, 
and B.\ Grossman, ibid B\u{203}, 205 (1982). Also see
K.\ Huang in Quarks, Leptons and Gauge Fields (World Scientific,
Singapore, 1992).

\item %10
\ See e.q. V.\ Dmitrasinovic, H.\ Schulze, R.\ Tegen and R.\ Lemmer,
Phys.\ Rev.\ D\u{52}, 2855 (1995). Combine their equs. (8) and (15) to
obtain our gap equation (6).

\item %11
A.\ Salam, Nuovo Cimento \u{25}, 224 (1962); S.\ Weinberg, Phys.\ Rev.\ 
\u{130}, 776 (1963).

\item %12
S. \ Weinberg, Phys.\ Rev.\ Lett. \u{17}, 616 (1966).

\item %13
K.\ Akama, Phys.\ Rev.\  Lett.\u{76}, 184 (1996) and J.\ Zinn Justin, 
Nucl. \ Phys.\ B\u{367}, 105 (1991) consider a four-quark 
$U(1)_{L}\times\ U(1)_R$ simple NJL model finding $\lambda = g^2$ and 
$m_{\sigma} \to 2m_q$ when $N_{c}\rightarrow\infty$.  But this does not 
correspond to the SU(2) \lsm\ with $m_{\sigma}=2m_q$ when $N_c=3$.

\item %14
See e.g.\ L.\ Dolan and R.\ Jackiw, Phys.\ Rev.\ D\u{9},
3320 (1974).

\item %15
D.\ Bailin, J.\ Cleymans and M.\ D.\ Scadron, Phys.\ Rev.\ D\u{31},
164 (1985); J.\ Cleymans, A.\ Koci\'c and M.\ D.\ Scadron, ibid D\u{39},
323 (1989).

\item %16
See e.g.\ review by T.\ Hasuda, Nucl.\ Phys.\ A\u{544}, 27 (1992);
also see M.\ Asaka and K.\ Yazaki, ibid.\ A\u{504}, 668 (1989);
A.\ Barducci, R.\ Casalfuoni, Phys.\ Rev.\ D\u{41}, 1610 (1990).

\item %17
N.\ Bili\'c, J.\ Cleymans and M.\ D.\ Scadron, Int.\ J.\ Mod.\ Phys.\
A\u{10}, 1169 (1995).

\item %18
See e.g.\ reviews by B.\ Petersson, Nucl.\ Phys.\ (Proc.\ Supp)
B\u{30}, 66 (1993); K.\ M.\ Bitar et al.\ ibid.\ B\u{30}, 315 (1993).

\item %19
N.\ Tornqvist and M.\ Roos, Phys.\ Rev.\ Lett.\ \u{76}, 1575 (1996).  

\item %20 
K.\ Rajagopal and F.\ Wilczek, Nucl.\ Phys.\ B \u{404}, 577
(1993); M.\ Asakawa, Z.\ Huang and X.\ -N.\ Wang, Phys.\ Rev.\ Lett.\ 
\u{74}, 3126 (1995); J.\ Randrup, ibid.\ \u{77}, 1226 (1996).

\item %21
L.\ P.\ Csernai and I.\ N.\ Mishustin, Phys.\ Rev.\ Lett.\ \u{74}, 5005
(1995).

\end{enumerate}
\end{document}